\documentclass[english,aps, two column]{revtex4}
\usepackage[T1]{fontenc}
\usepackage[latin9]{inputenc}
\usepackage{color}
\usepackage{textcomp}
\usepackage{amsmath}
\usepackage{amssymb}
\usepackage{graphicx}
\usepackage{esint}

\makeatletter

\providecommand{\tabularnewline}{\\}
\@ifundefined{textcolor}{}
{%
 \definecolor{BLACK}{gray}{0}
 \definecolor{WHITE}{gray}{1}
 \definecolor{RED}{rgb}{1,0,0}
 \definecolor{GREEN}{rgb}{0,1,0}
 \definecolor{BLUE}{rgb}{0,0,1}
 \definecolor{CYAN}{cmyk}{1,0,0,0}
 \definecolor{MAGENTA}{cmyk}{0,1,0,0}
 \definecolor{YELLOW}{cmyk}{0,0,1,0}
 }

\makeatother

\usepackage{babel}
\begin{document}


\title{High repetition rate and coherent Free-Electron Laser in the tender
	X-rays based on the Echo-Enabled Harmonic Generation of an Ultra-Violet
	Oscillator pulse}

\author{N. S. Mirian$^{1}$, M. Opromolla$^{2,3}$, G. Rossi$^{2}$, L. Serafini$^{3}$
	and V.~Petrillo$^{2,3}$}

\affiliation{$^{1}$Deutsches Elektronen-Synchrotron DESY, 22607 Hamburg, Germany}

\affiliation{$^{2}$Universit� degli Studi di Milano,Via Celoria,16 20133 Milano,
	Italy}

\affiliation{$^{3}$INFN - Sezione di Milano, Via Celoria 16, 20133 Milano and
	LASA, Via F. Cervi 201, 20090 Segrate (MI), Italy}

\email{najmeh.mirian@desy.de}
\email{Michele.Opromolla@mi.infn.it}

\selectlanguage{english}%
\begin{abstract}
	Fine time-resolved analysis of matter - i.e. spectroscopy and photon
	scattering - in the linear response regime requires a fs-scale pulsed,
	high repetition rate, fully coherent X-ray source. A seeded Free-Electron
	Laser (FEL) driven by a Super-Conducting Linac,
	generating $10^{8}$-$10^{10}$ coherent photons at 2-5 keV with about
	0.5 MHz of repetition rate, can address this need. The seeding scheme
	proposed is the Echo-Enabled Harmonic Generation, alimented by a FEL
	Oscillator working at 13.6 nm with a cavity based on Mo-Si mirrors.
	The whole chain of the X-ray generation is here described by means
	of start-to-end simulations. Comparisons with the Self Amplified Spontaneus
	Emission and a fresh-bunch harmonic
	cascade performed with similar electron beams show the validity
	of this scheme.
\end{abstract}
\maketitle

\section{Introduction}

Fine time-resolved analysis of matter is currently performed with synchrotron
radiation (SR) sources or with X-ray Free Electron Lasers (FELs){, whose extremely brilliant pulses are able to detect matter in highly excited states dominated by non-linear response.}
Spectroscopic studies and photoemission experiments require probes with fluxes smaller than 10$^{8}$ photons/pulse and MHz-class repetition rates, for remaining below the linear response threshold and collecting adequate statistics. Current FELs' photon number exceeds this level by 2-4 orders of magnitude, requiring severe attenuation with huge waste of energy. On one hand, sources based on Warm
Linacs, operating at 10/120 Hz, are inadequate for collecting statistics for high resolution spectroscopy.
On the other hand, signals like the one by EuXFEL \cite{XFEL,xfel2}, shaped in thousands micropulses grouped in 10 macropulses per second, are also non ideal for spectroscopy as both attenuation is needed and the high repetition rate of the micropulses overruns detector and pump-probe set-up capability.
%The extremely brilliant FEL flashes, down to the X-range, shorter than 100 fs and with more than $10^{12}$ photons/shot, are able to detect matter in highly excited states dominated by non-linear response and test individual objects before destruction.
%in the self-amplified spontaneous emission (SASE) mode \cite{flash,LCLS,SACLA,XFEL,xfel2, PSI,PAL}.
%FERMI@Elettra \cite{fermi2} is the unique fully coherent operating FEL in the Ultra-Violet (UV) and soft X-ray regimes
%but studies on seeding techniques are ongoing worldwide.   
{SASE fluctuations severely limit the use of FELs in X-ray spectroscopy and full seeding, \textcolor{black}{routinely implemented at FERMI@Elettra \cite{fermi2} and SXFEL \cite{selfmod}, should be extended to tender/hard X-ray energies.}}
	%in the Ultra-Violet (UV) and soft X-ray regimes

There is therefore scientific need and ample room for a novel type
of source: a source delivering to the sample $10^{7}-10^{8}$ photons
in 10 fs coherent pulses at 0.5-2 MHz in the tender/hard X-ray range,
thus bridging the gap in time resolution and average photon flux between
the most advanced SR and the current FELs. These requests are addressed by conceiving a tailored seeded FEL
driven by Linacs based on Super-Conducting cavities, providing $10^{8}$-$10^{10}$
coherent photons at 2-5 keV, at about 1 MHz of repetition rate. \\
In the seeded FEL configuration, an external coherent pulse imprints its temporal phase on the electron beam at the undulator entrance.
The direct seeding \cite{directseeding} is not possible in the soft-hard
X-ray range due to the lack of high-power coherent seeds at these
wavelengths, while self-seeding processes \cite{self1,self2} only achieve partial coherence. 
%Novel sources of high harmonics in gas are now capable of producing short ($< 100$ fs) XUV pulses with high energy resolution and a repetition rate of about 200 kHz \cite{HHGnew}.
High Gain Harmonic Generation (HGHG) multistage cascades \cite{lee}, \textcolor{black}{seeded by the harmonics
of an IR laser generated in crystals \cite{dcls} or in gases \cite{HHG, hhg2,Sparc_cascata,sparc in gas}}, have been demonstrated and applied up to few nm wavelengths \cite{fermi2}. 
However, their implementation in the tender/hard X-ray range is highly
demanding, while the extention to higher repetition rates, obtained
by using oscillators \cite{NJP_Oscill} or lasers in cavity \cite{desy},
has been studied sofar only theoretically. 
{As demonstrated at SXFEL \cite{selfmod}, a step towards high repetition rate seeded FELs is also foreseen by using an optical klystron-like configuration, allowing to reduce the requirement for the peak power of the seed laser.}
%have been pioneeringly studied in the optical-UV range \cite{taga, taga2, Sparc_cascata,sparc in gas}, then  
%Operation in single spike SASE mode \cite{singlesp,singlesp2,singlsp3} or self-seeding processes \cite{self1,self2} achieve partial coherence. 
FEL oscillators \cite{oscill,dattoli-1,Freund,Xfelo2,xfelo3,instru} or regenerative amplifiers \cite{reg ampl-1,mcneil,lind,rafel_2019,Marcusconfronto} could directly produce coherent X-rays \cite{oscill seed-1,NJP_Oscill}, but the operational scenario proposed so far, with electrons at several GeVs, impedes their
realization in small/medium size research laboratories. UV/soft X-ray coherent radiation has been generated with the Echo-Enabled Harmonic
Generation (EEHG) \cite{EEHG0,EEHG,EEHG-1,EEHG-2,EEHG-3}, a technique which requires two coherent radiation pulses, usually delivered by optical lasers, seeding the electron beam in two sequential modulators
interspersed by a strong dispersive section. 
A second dispersive section and the radiator are placed dowstream. As explained
in Ref.s \cite{EEHG0,EEHG}, the combination of energy modulation and dispersion, replicated twice, warps the electron longitudinal phase space, producing a significant bunching on very high harmonics which drives the emission of a short wavelength coherent pulse in the radiator. 
{A combined EEHG-oscillator scheme with two oscillators as modulators has also been proposed \cite{eehg-osc}.}
%, suffering from not-controllable energy modulations, 

\begin{table}
	\caption{\label{tab:Electron-beam-and-1}Electron beam for MariX FEL. }
	\begin{tabular}{|c|c|c|}
		\hline 
		electron beam energy & GeV & 2-3.8\tabularnewline
		\hline 
		Charge & pC & 30-50\tabularnewline
		\hline 
		Current & kA & 1.6\tabularnewline
		\hline 
		rms normalized emittance & mm mrad & 0.3-0.5\tabularnewline
		\hline 
		rms relative energy spread & $10^{-4}$ & 3\tabularnewline
		\hline 
		electron beam duration & fs & 12\tabularnewline
		\hline 
		slice energy spread & $10^{-4}$ & 4-2\tabularnewline
		\hline 
		slice normalized emittance & mm mrad & 0.3\tabularnewline
		\hline 
	\end{tabular}
\end{table}
In this paper, we show the operation of a FEL in the tender X-ray
range, based on EEHG. 
%Differently from all previous experiments and studies, 
{We propose a FEL Oscillator in the far Ultra-Violet
frequency range as seeding source}. The advantage of such a scheme is twofold: the oscillator
operation, combined with an electron beam accelerated in a Super-Conducting
Linac, increases the device repetition rate by many orders
of magnitude (4-6). Moreover, since the oscillator frequency and peak power are much
higher than the ones of a conventional laser, the soft-hard X-ray range
can be more easily reached by a lower harmonic number.

\section{Lay-out of the coherent source and simulations}

The electron beam is supposed
to be generated by an accelerator similar to the project MariX's (Multi-disciplinary
Advanced Research Infrastructure for the generation and application
of X-rays) \cite{mariX1-1,bacciprab-1}, whose compact footprint with a total length of about 500 m and contained costs should permit its
construction also in medium-size research infrastructures or within
university campuses. MariX is based on the innovative design of a
two-pass two-way Super-Conducting linear electron accelerator \cite{bacciprab-1},
equipped with an arc compressor \cite{arco-1,arco2-1} to be operated  in CW mode at 1 MHz. The characteristics of the electron beam are listed
in Table \ref{tab:Electron-beam-and-1}.
\begin{figure*}[!ht]
	\includegraphics[width=1\textwidth]{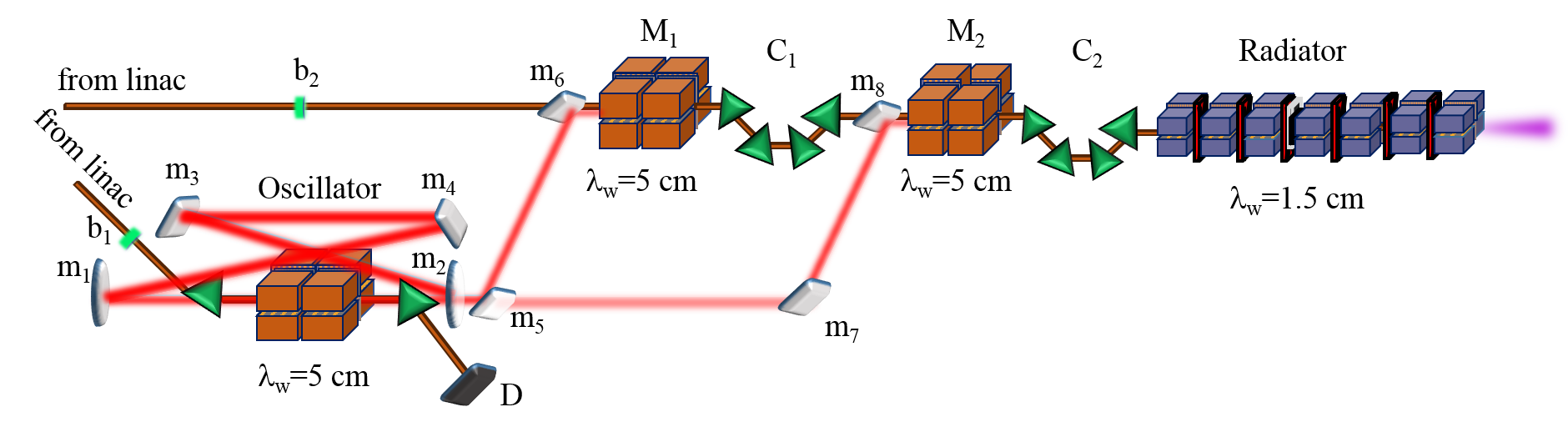}\caption{\label{fig:Regenerative-Amplifier-with}Scheme of the Echo-Enhanced
		Harmonic Generation FEL. $\mathrm{b}_{1}$ and
		$\mathrm{b}_{2}$: electron bunches alternatively sent in the oscillator and in
		the undulators.  FEL Oscillator undulator: $\lambda_{w}$=5 cm, $a_{w}$=3.71,
		total length 9 m.
		$m_{1}$, $m_{2}$, $m_{3}$, $m_{4}$: oscillator cavity mirrors.
		D: beam dump. $m_{5}$,
		$m_{6}$, $m_{7}$, $m_{8}$: mirrors of the optical line that splits and couples the seed to the
		modulators. $m_{2}$ and $m_{5}$: beam splitters. First ($M_{1}$)
		and second ($M_{2}$) modulators: $\lambda_{w}$=5 cm, $a_{w}$=3.71, respectively 1.6 m and 1.5 m long. The
		two chicanes $C_{1,2}$ are both 0.45 m long. Radiator:
		about 20 m long,  $\lambda_{w}$=1.5 cm, maximum magnetic field: about 1 T.}
\end{figure*}
{As studied and demonstrated in Ref. \cite{hom}, the superconducting technology allows to achieve low jitters and
fluctuations of the electron beam. These increased stability conditions, together with the seeded mode FEL
operation, give the possibility to produce a fully coherent, high repetition rate and highly stable x-ray source.}

Fig. \ref{fig:Regenerative-Amplifier-with} shows the scheme of the
source. After the acceleration stage, successive electron bunches
are alternatively driven in the oscillator or matched to the EEHG
undulator device. {The electron beam alimenting the oscillator is extracted upstream the linac end at an energy of 2 GeV, while the ones entering the EEHG device may have the same energy or be further accelerated.} 
%the same energy or extracted upstream the linac end,
The oscillator is constituted
by a 9 m long undulator segment with period $\lambda_{w}=5$ cm and
produces 70 $\mu J$ of intracavity radiation at $\lambda_{O}=$13.6
nm. It is embedded into a folded ring cavity composed
by 4 mirrors, two of which focusing{, with optics heat loading requiring an intelligent cooling system.} For an oscillator repetition rate
of 0.5 MHz, the round trip length $L_{c}$ is 600 m and the distance
between two mirrors is $L_{c}/4=150$ m. The oscillator supermodes
\cite{oscill} are calculated fully numerically \cite{mcneil,NJP_Oscill}
by extracting the radiation field simulated by {\small GENESIS} {\small 1.3
	\cite{genesis-1}} from the oscillator, driving it through the optical
line accounting for mirrors and propagation, and superimposing it
on the successive electron bunch. The microscopic distribution of the
electron beam is changed shot to shot in order to simulate the passage
of a sequence of different bunches. After the
passage into the oscillator, the electron bunch is deteriorated by the
radiation process and driven to the dump. 
%OFEL requires the intelligent cooling system to load the heating.

\begin{figure}[!ht] \includegraphics[width=0.9\columnwidth]{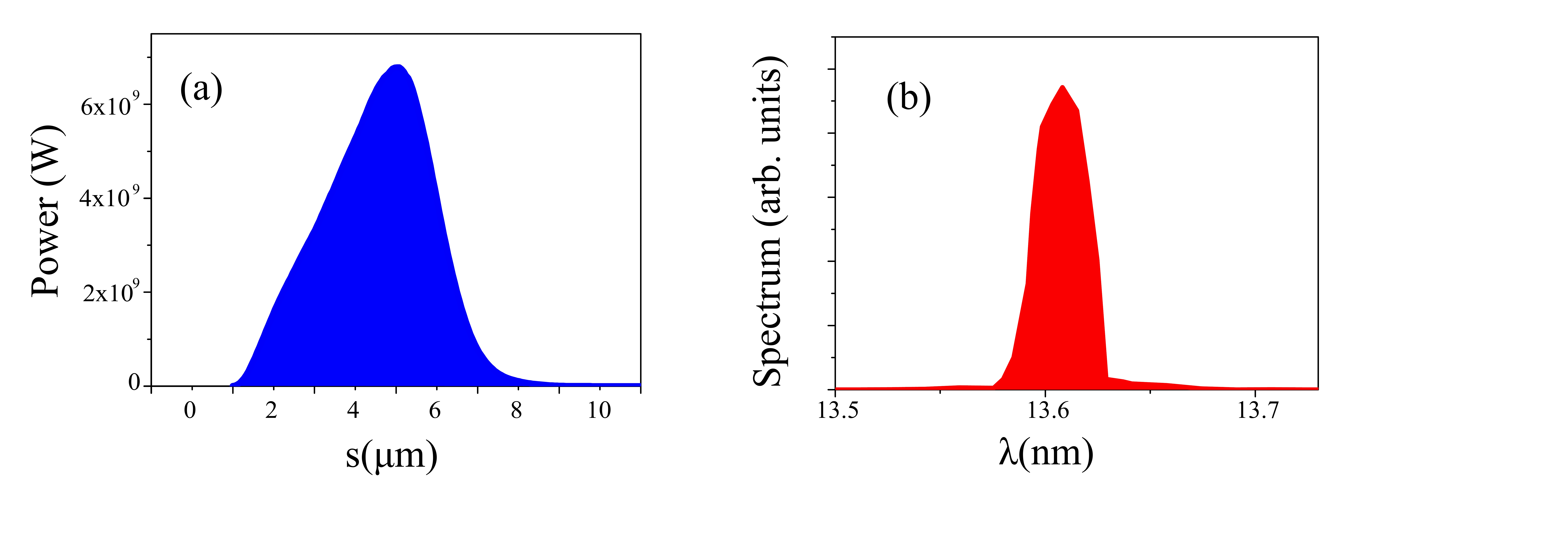}\caption{\label{fig:Signal-and-spectrum}Signal (a) and spectrum (b) of the intracavity power of the oscillator at saturation.} \end{figure}

Fig. \ref{fig:Signal-and-spectrum} presents the intracavity pulse temporal and spectral densities of the seed at saturation, whose 
Table \ref{tab:Characteristics-of-the} summarizes the characteristics of the intracavity seed pulse at saturation.
After an optical transport line that splits it in two pulses, the
seed is synchronized to the electron bunches at the beginning of both
modulators. Energy losses along the transport line have been taken into account. 
{The oscillator seed pointing stability and
transverse overlap with the electron bunches after the transport line will be checked and adjusted with
multipurpose stations and beam stabilization systems as the ones described in \cite{end_stat1,end_stat2}}.
\begin{table}
	\caption{\label{tab:Characteristics-of-the}Characteristics of the seed, generated
		by the FEL oscillator with undulator length $L_{w}$=9 m. The repetition
		rate of the source is 0.5 MHz. \$=Photons/s/$mm^{2}$/$mrad^{2}$/bw(\textperthousand{}).}
	\begin{tabular}{|c|c|c|c|}
		\hline 
		$\lambda_{O}$(nm) & 13.6  & E($\mu$J) & 70\tabularnewline
		\hline 
		Photons/shot & 4.5$\times10^{12}$ & Photons/s & 2.27$\times10^{18}$\tabularnewline
		\hline 
		relative bandwidth & 1.7$\times10^{-3}$ & rms length($\mu m$) & 2\tabularnewline
		\hline 
		div($\mu rad$) & 50 & size($\mu$m) & 80\tabularnewline
		\hline 
		peak brilliance(\$) & 2.4$\times10^{31}$ & average brilliance(\$) & 8$\times1$$0^{22}$\tabularnewline
		\hline 
	\end{tabular}
\end{table}

\begin{figure*}[!ht]
	\includegraphics[width=0.9\textwidth]{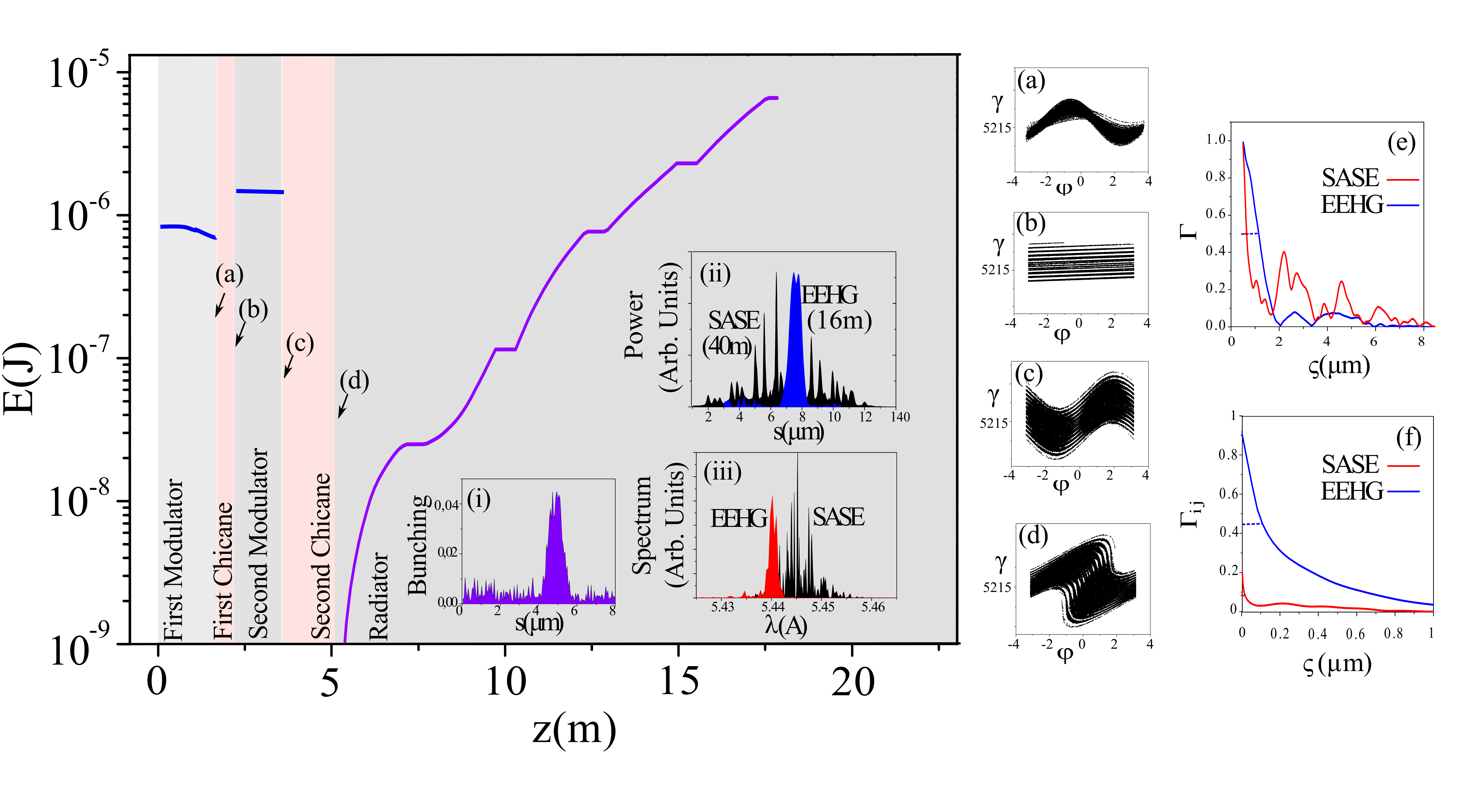}
	\caption{\label{fig:Growth-25-H.}Radiation energy E(J) vs z(m). In blue, radiation
		at $\lambda_{o}=$13.6 nm in the modulators. In violet, radiation
		at $\lambda=$13.6/25 nm=5.44 \mbox{\normalfont\AA} in the radiator. Point (a) is the
		$1^{st}$ modulator end, (b)  $1^{st}$ chicane end, (c)
		$2^{nd}$ modulator end and (d) $2^{nd}$ chicane end.
		Inner figures: (i) bunching on the $25^{th}$ harmonics
		at the end of the $2^{nd}$ chicane, (ii)  power density, (iii)  spectral amplitude (SASE at 40 m and EEHG at the radiator exit). Windows (a), (b), (c), (d): electron
		phase space, electron Lorentz factor $\gamma$ vs seed phase $\varphi$, in the corresponding points. (e) and (f): SASE (red) and EEHG (blue) coherence degree for one pulse $\Gamma$
		and for different pulses $\Gamma_{ij}$ vs $\varsigma$=c$\tau$. Dotted segments: FWHM
		coherence lengths.}
\end{figure*}

To obtain radiation in the range $\lambda=$5-2 \mbox{\normalfont\AA}, namely 2-5 keV,
with the MariX's moderate energy electron bunch (2.5-3.8 GeV), a short
period radiator must be considered. From the resonance relation $\lambda=\frac{\lambda_{w}}{2\gamma^{2}}(1+a_{w}^{2})$
($a_{w}$ is the undulator parameter and $\gamma$ the electron Lorentz
factor), taking a maximum magnetic field B=1T corresponding to $a_{w}$=0.98,
we can deduce that an undulator period of $\lambda_{w}$=1.5 cm is
suitable. Besides, starting from a seed at $\lambda_{o}=$
13.6 nm, a significant electron bunching on harmonics'
order n in the range from 25 to 70 is required. By following the uni-dimensional
model based on plane waves exposed in Ref.s \cite{EEHG0,EEHG}, the
electron bunching is expressed in terms of four free parameters, namely
the dispersion strengths of the two chicanes $R_{56,1}$ and $R_{56,2}$
and the normalized electric fields of the seeds: 
\begin{equation*}
A_{1,2}=\frac{ea_{w}\lambda_{O}}{mc^{2}\pi\sigma_{1,2}}\sqrt{\frac{P_{1,2}}{c\varepsilon_{0}\pi}},
\end{equation*}
$P_{1,2}$ and $\sigma_{1,2}$ being the peak power and the rms transverse
dimension of the seeds, e electron charge, m electron mass, c speed
of light and $\varepsilon_{0}$ the vacuum dielectric constant. Assuming
$P_{1,2}=100$,200 MW respectively (0.86 and 1.72 $\mu J$ of total
energy), and keeping the longitudinal lengths of the two chicanes
constant, $R_{56,1}$ and $R_{56,2}$ have been optimized
to maximize the electron bunching on the desired harmonics
of the seed. Using an electron energy of 2.66 GeV and a slice relative
energy spread of $\Delta E/E=3\times10^{-4}$, proper
bunching at the second chicane end is obtained for n=25 , corresponding
to $\lambda=5.44$ \mbox{\normalfont\AA} with $R_{56,1}$=132 $ \mu m$ and $R_{56,2}$=4.72 $\mu m$. The power growth in this case is shown in Fig.
\ref{fig:Growth-25-H.}, its central windows showing the electron phase spaces after the first
modulator (a), after the first chicane (b), after the second modulator
(c) and at the radiator entrance (d). The initial bunching on $\lambda=$5.44 $\mbox{\normalfont\AA}$ (in violet) reaches a peak of 4.5\% on the bunch, while the rms energy
spread increases from 0.03\% to 0.045\% in the modulators and in the
chicanes. These parameters are sufficient to trigger a consistent
FEL emission in the radiator, and the radiation is extracted at the minimum
bandwidth position, occurring after 12 m of radiator (about 16 m of
the total device). The neat single spike structures in power and spectral
amplitude at the end of the undulator,
are shown in the inner windows in blue and red respectively, compared with the corresponding SASE
profiles extracted after 40 m of undulator. Source coherence and stability
are evaluated through the modulus of the complex coherence degree:
\begin{equation*}
\Gamma_{ij}(\tau)=\left|\frac{\int dtE_{i}(t)E_{j}(t-\tau)}{\sqrt{\int dt\left|E_{i}\right|^{2}}\sqrt{\int dt\left|E_{j}\right|^{2}}}\right|,\label{eq:2}
\end{equation*}
between two different generic pulses i and j generated by different
electron beams ($\Gamma_{ij}$) or for one single pulse ($\Gamma=\Gamma_{i,i}$),
where $E_{i}$ and $E_{j}$ are the electric fields of the pulses
as function of time.
%$\Gamma=\Gamma_{ii}$ and $\Gamma_{ij}$ 
These two quantities are shown vs $\varsigma=c\tau$
for n=25 in the right windows of Fig. \ref{fig:Growth-25-H.}, together
with those of the SASE case at saturation. The FWHM coherence length of
the EEHG pulse is $\mathfrak{L}_{c}=0.67\,\mu m$, about 4 times the
corresponding SASE's. The equal time coherence between different pulses
$\Gamma_{ij}(0)$ is close to 1, differently from the SASE's of
$10^{-1}$. The FWHM mutual coherence length, quantifying the shot-to-shot stability, is 0.1 $\mu m$, compared to 0.015 $\mu m$ of the SASE
one. The EEHG coherent source provides an ultrashort pulse with 5
$\mu J$ of energy per shot at $\lambda=$5.44 \mbox{\normalfont\AA}, corresponding to
$1.4\mathrm{\times}10^{10}$ photon/shot at 0.5 MHz or $0.7\times10^{16}$
photons/s, while the SASE source reaches $14.7\times10^{16}$ photons/s
at 1 MHz. Even if characterized by a smaller number of photons, bandwidth, coherence and collimation of the EEHG give an average brilliance
of $1.7\times10^{24}$ photons/s/$mm^{2}$/$mrad^{2}$/bw(\textperthousand{}), being only a factor 5 smaller than the corresponding SASE's. 
\begin{figure}[!ht]
	\includegraphics[width=0.9\columnwidth]{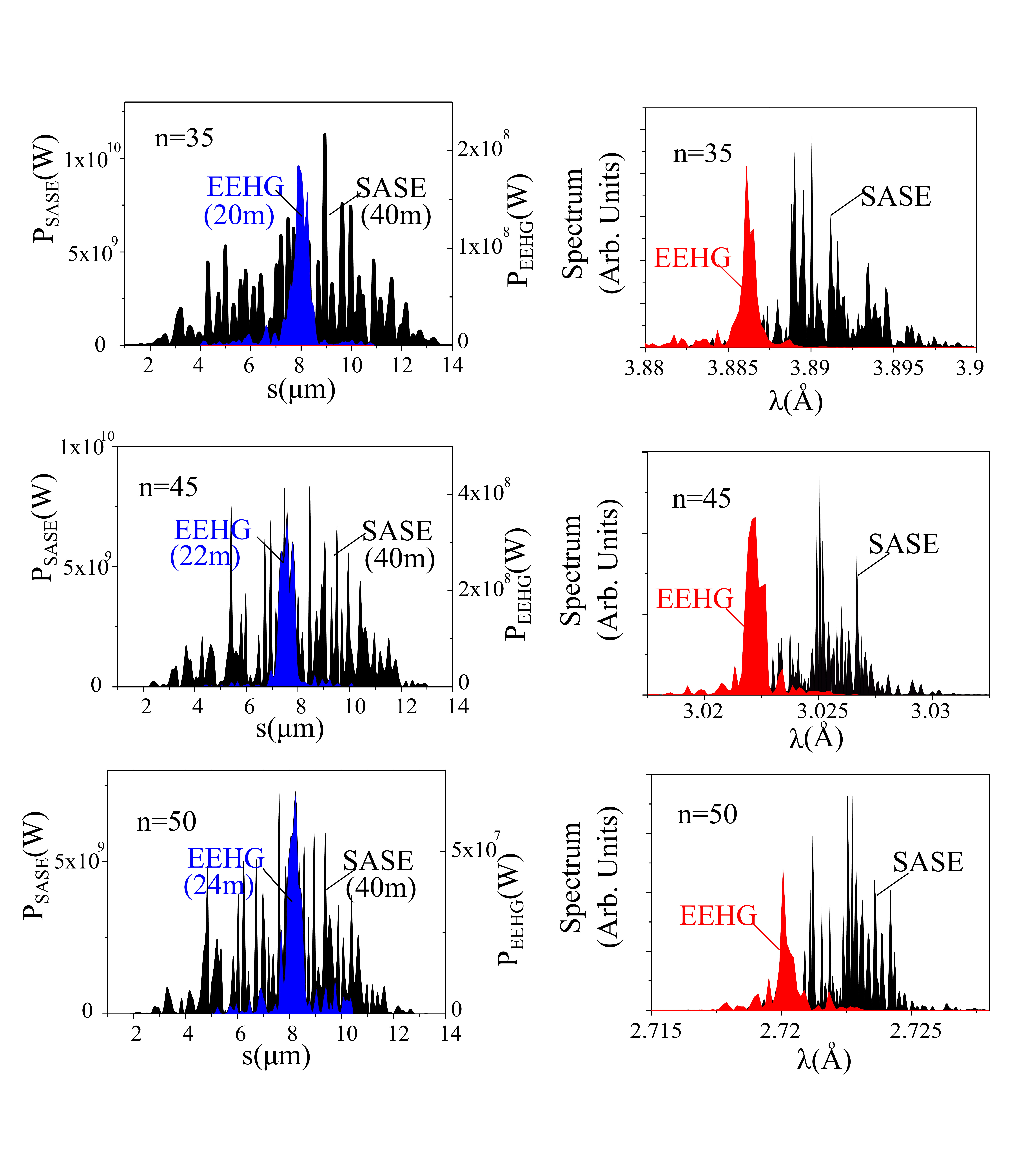}
	\caption{\label{fig:Spettri}SASE (black) and EEHG (red) power and spectral radiation profile for n=35 ($R_{56,1}$=184.7$\mu m$, $R_{56,2}$=4.86$\mu m$), 45 ($R_{56,1}$=281.6 $\mu m$, $R_{56,2}$=6.58$\mu m$) and 50 ($R_{56,1}$=293.2$\mu m$,$R_{56,2}$=6,07 $\mu m$).  SASE power vs s=ct is on the left axis,  EEHG power on the
		right. Spectra are in arbitrary units and not in scale.}
\end{figure}
\begin{table*}
	\caption{\label{tab:MariX-FEL-characteristics}SASE (extracted at saturation), EEHG (extracted at minimum bandwidth) and fresh-bunch HGHG cascaded FEL characteristics, \$=Photons/s/$mm^{2}$/$mrad^{2}$/bw(\textperthousand{}).}
	\begin{tabular}{|c|c|c|c|c|c|c|c|c|c|c|c|}
		\hline 
		Mode &  & SASE & EEHG & HGHG & SASE & EEHG & HGHG & SASE & EEHG & SASE & EEHG\tabularnewline
		\hline 
		Harmonic number &  &  & 25 & 5$\times$5 &  & 35 & 7$\times$5 &  & 45 &  & 50\tabularnewline
		\hline 
		Wavelength & \mbox{\normalfont\AA} & 5.44 & 5.44 & 5.44 & 3.88 & 3.88 & 3.88 & 3.02 & 3.02 & 2.72 & 2.72\tabularnewline
		\hline 
		
		$\gamma$ &  & 5216 & 5216 & 5216 & 6151 & 6151 & 6151 & 6991 & 6991 & 7352 & 7352\tabularnewline
		\hline 
		Undulator length & m & 40 & 16 & 32 & 40 & 20 & 32 & 40 & 22 & 40 & 24\tabularnewline
		\hline 
		\hline

		Energy & $\mu J$ & 58 & 5.1 & 7 & 58 & 0.9 & 3.2 & 46 & 0.9 & 42 & 0.18\tabularnewline
		\hline
		Photon/shot & $10^{9}$ & 158 & 13.9 & 19 & 112 & 1.74 & 6.2 & 69 & 1.36 & 57 & 0.245\tabularnewline
		\hline 
		
		Bandwidth & 0.1\% & \textcolor{black}{1} & 0.37 & 0.17 & 0.9 & 0.4 & 0.32 & 0.6 & 0.3 & 0.5 & 0.35\tabularnewline
		\hline 
		Length & fs & \textcolor{black}{10} & 1.5 & 3 & 10 & 0.8 & 4.5 & 10 & 1 & 10 & 1\tabularnewline
		\hline 
		Divergence & $\mathrm{\mu rad}$ & \textcolor{black}{3.7} & 3.9 & 3.6 & 2.7 & 3.2 & 5.1 & 2.2 & 2.5 & 2.05 & 2.3\tabularnewline
		\hline 
		Pulse size & $\mathrm{\mu m}$ & 35 & 27 & 24 & 32 & 25 & 29 & 27 & 25 & 27 & 30\tabularnewline
		\hline 
		Photon/s & $10^{15}$ & 158 & 7 & 4.7 & 112 & 0.87 & 3.1 & 69 & 0.67 & 57 & 0.09\tabularnewline
		\hline 
		Average brilliance & $10^{23}$ \$ & 94 & 17 & 37 & 151 & 3.39 & 5.51 & 325 & 5.71 & 251 & 0.51\tabularnewline
		\hline 
		Coherence length & $\mu m$ & 0.17 & 0.67 & 1.3 & 0.11 & 0.52 & 1 & 0.1 & 0.5 & 0.1 & 0.3\tabularnewline
		\hline 
	\end{tabular} 
\end{table*}
Working points covering the range between 5.44 and 2.7 \mbox{\normalfont\AA}, namely n=25,
35, 45 and 50, are presented in Table \ref{tab:MariX-FEL-characteristics}
and compared to similar SASE cases after 40 m of undulator. Electron
beams with progressively larger energies  allow to reach shorter wavelengths. Fig. \ref{fig:Spettri}
presents powers and spectral amplitudes for both SASE (black) and
EEHG (red) cases for n=35, 45, 50. The case with n= 35 is performed with an electron beam of about 3 GeV and delivers 1.7$\times10^{9}$ photons/shot at 3.88 \mbox{\normalfont\AA}, extracted after 16 m of radiator
with an average brilliance of $3.4\times10^{23}$ photons/s/$mm^{2}$/$mrad^{2}$/bw(\textperthousand{}). 
With an energy of 3.6 GeV, 1.36$\times10^{9}$ photons/shot at 3.02 \mbox{\normalfont\AA} (n=45) can be generated
with an average brilliance of 5.7$\times10^{23}$ photons/s/$mm^{2}$/$mrad^{2}$/bw(\textperthousand{}). 
Pushing the electron energy to the maximum value foreseen for MariX, 3.8 GeV, the system produces 2.4$\times10^{8}$ photons /shot at 2.7 \mbox{\normalfont\AA} (n=50) with a brilliance of 5.7$\times10^{23}$ photons/s/$mm^{2}$/$mrad^{2}$/bw(\textperthousand{}).
The coherence length is from 5 (n=35) to 3 (n=50) times the corresponding
SASE's, while the stability is larger by a factor from 10 (n=35) to
20 (n=50). 
Since these estimations widely exceed the target values
set by the scientific case, the EEHG source will be capable to satisfy
the conditions requested by the envisaged experiments, considering
a safety margin compensating all the degradations due to errors, misalignment,
jitters and the losses dealing with the transport of the photon beams
to the experimental hutch. The comparison between the results of the
EEHG technique and those of a fresh-bunch HGHG cascade, performed
with a similar electron beam and with the same oscillator as seeding
source \cite{NJP_Oscill}, shows a substantial equivalence in terms
of number of emitted photons for n=5$\times$5 and a better performance
of the HGHG cascade for n=7$\times$5, when the induced energy
spread limits the EEHG radiation. However, the HGHG efficiency decreases
very rapidly at higher harmonics (n>35) and radiation levels comparable
with the EEHG's cannot be achieved. A total of four different electron
bunches per shot are needed for the fresh-bunch technique, while
for the EEHG only one bunch for the oscillator and one for the FEL
are required, doubling, in proportion, the radiation repetition rate.
Regarding coherence, a better performance of the HGHG cascade is observed,
due to the more direct transfer of the coherence properties from the
seed to the radiation. The advantage of the EEHG is a larger tunability
and versatility of the source, that permits to generate intense ($10^{10}-10^{8}$
photons/shot), ultra-short (down to 1 fs) pulses at all the harmonics
n of the seed up to n=50 and not only at those corresponding to the
product of two odd integer numbers (n=5$\times$5, 3$\times$9, 7$\times$5...). 

\section{Conclusions}
A new generation accelerator complex is at the core of this coherent
and compact facility dedicated and optimized to ultra-fast coherent
X-ray spectroscopy and inelastic photon scattering, and to highly
penetrating X-ray imaging of mesoscopic and macroscopic samples.
\textcolor{black}{The X-ray generation scheme here studied relies on a conventional EEHG device seeded by a far-UV FEL Oscillator and reaching high harmonic orders up to n=50.
Its comparison with a SASE FEL performed with a similar electron beam proves the much higher stability and coherence of the produced pulses. The major advantage with respect to a fresh-bunch three-stage FEL cascade seeded by the same oscillator is given by the tunability and simpler set-up of the EEHG scheme.}
%The capability to generate stable, coherent and high repetition rate tender X-ray FEL pulses.
%This source will permit the XAS/XMCD (with polarization control from quarterwavelength blades or undulators) and bulk photoemission applications and could become a highly efficient probe of matter at the nanoscale in bulk environments as, for instance, buried interfaces of interest in materials science, in-vivo biological samples or catalysers at work.
Such facility will be intrinsically multi-user and multidisciplinary as
of the research performed and science output.
%It will therefore create absolutely novel conditions for experiments that cannot be performed satisfactorily at the present and foreseen sources based on storage rings or SASE-FELs, and, due to its compact footprint, could be implemented in medium sized research laboratories and universitary campuses. The anticipated performances of MariX Free Electron Laser are well beyond the state of the art of most of presently FELs in operation, and in the trailing edge of EuXFEL and of the US future superconducting FEL project of reference, LCLS-II.

\end{document}